\documentclass{article}
\usepackage{spconf,amsmath,graphicx}
\usepackage{multirow}
\usepackage{multicol}
\usepackage{lscape}
\usepackage{booktabs}
\usepackage{amsfonts, eucal}

\title{nnSpeech: Speaker-Guided Conditional Variational Autoencoder for Zero-shot Multi-speaker Text-to-Speech}

\name{Botao Zhao$^{1,2}$, Xulong Zhang$^{1}$, Jianzong Wang$^{1*}$, Ning Cheng$^{1}$, Jing Xiao$^{1}$
\thanks{$^*$ Corresponding author: Jianzong Wang, jzwang@188.com}}
\address{$^{1} $Ping An Technology (Shenzhen) Co., Ltd., China \\ $^{2} $University of Auckland, New Zealand}
\address{
  $^1$Ping An Technology (Shenzhen) Co., Ltd., China\\
  $^2$Institute of Science and Technology for Brain-inspired Intelligence, Fudan University, China}

\begin{document}
%
\maketitle
\begin{abstract}
Multi-speaker text-to-speech (TTS) using a few adaption data is a challenge in practical applications. To address that, we propose a zero-shot multi-speaker TTS, named nnSpeech, that could synthesis a new speaker voice without fine-tuning and using only one adaption utterance. Compared with using a speaker representation module to extract the characteristics of new speakers, our method bases on a speaker-guided conditional variational autoencoder and can generate a variable Z, which contains both speaker characteristics and content information. The latent variable Z distribution is approximated by another variable conditioned on reference mel-spectrogram and phoneme. Experiments on the English corpus, Mandarin corpus, and cross-dataset proves that our model could generate natural and similar speech with only one adaption speech.
\end{abstract}

\section{Introduction}
\label{sec:intro}
Recently, with the great success of neural network based text-to-speech~(TTS) techniques, such as Tacotron~\cite{wang2017tacotron}, FastSpeech 1/2/s~\cite{ren2019fastspeech,ren2020fastspeech}, and Hifi-GAN~\cite{kong2020hifi} like neural vocoder, we can efficiently synthesize more natural voice. The single-speaker TTS model could be extended to multi-speaker scenarios using multi-speaker corpora~\cite{ren2019fastspeech,asru2021zhang}. However, these models are only applied to the fixed set of speakers and the training requires a certain amount of human recordings. 

Multi-speaker TTS for unseen speaker has many applications, such as news broadcast, personal assistant and metaverse. Several studies have been done in this field and one popular solution is using few data to fine-tune the base model that was trained by a large amount of data~\cite{cong2020data,chen2021adaspeech}. It achieves great voice quality when fine-tuning the whole model~\cite{kons2019high} or decoder~\cite{moss2020boffin}. There are also many studies only fine-tuning the speaker embedding~\cite{arik2018neural}. However, fine-tuning based methods have two distinctive challenges: 1) Fine-tuning the model means the memory storage and serving cost increasing because that we have to train many new models for each speaker. 2). Fine-tuning cost a huge amount of computing when we have a large number of users. 
Zero-shot multi-speaker TTS has become a popular field in recent years~\cite{wang2020bi,cooper2020zero,min2021meta}, which can synthesize voices directly by the trained base model. In some studies, pre-trained speaker verification models are used to extract speaker information~\cite{jia2018transfer}. We can also optimize the speaker encoder and TTS system jointly~\cite{chen2019cross} to make the speaker representation more optimal for the TTS system. But it is hard to extract the speaker representation to guide the TTS system and this could decrease voice quality or similarity. Zero-shot TTS needs training a generic model that could perform well on unseen data. However, due to the amount of data and the high dimension of the speaker vector, we can not estimate the distribution of speaker representation correctly. A solution is to use more adaption voices to get the averaged speaker representations~\cite{cooper2020zero}. But there are some applications requiring TTS model generate speech using a few adaption samples, such as personalized speech generation. 

To solve the above problem, we introduce conditional variational autoencoder~(CVAE) to this task. CVAE~\cite{sohn2015learning,zhao2017learning} is one of the most popular conditional directed graphical models whose input observations modulate the prior on Gaussian latent variables that generate the outputs. In this paper, we propose a new method using both CVAE and the prior knowledge~\cite{zhao2017learning}, speaker embeddings, for the zero-shot multi-speaker TTS. Our model is named of \textbf{nnSpeech}~(no new speech) because our method can synthesize new voices directly instead of fine-tuning a new model. In our TTS system, the phoneme is the condition. We assume that the distribution of latent variable $Z$ with the conditions of phoneme and speaker in multi-speaker TTS model could be approximated by the another variable conditioned on a reference mel-spectrogram and phoneme. To better training CVAE, we modified the standard CVAE and denotes it as speaker-guided CVAE~(sgCVAE).

\section{METHOD}
\label{sec:pagestyle}
\subsection{VAE and CVAE} 

\begin{figure}[htb] \label{fig: fig1}
  \centering
  \centerline{\includegraphics[width=7cm]{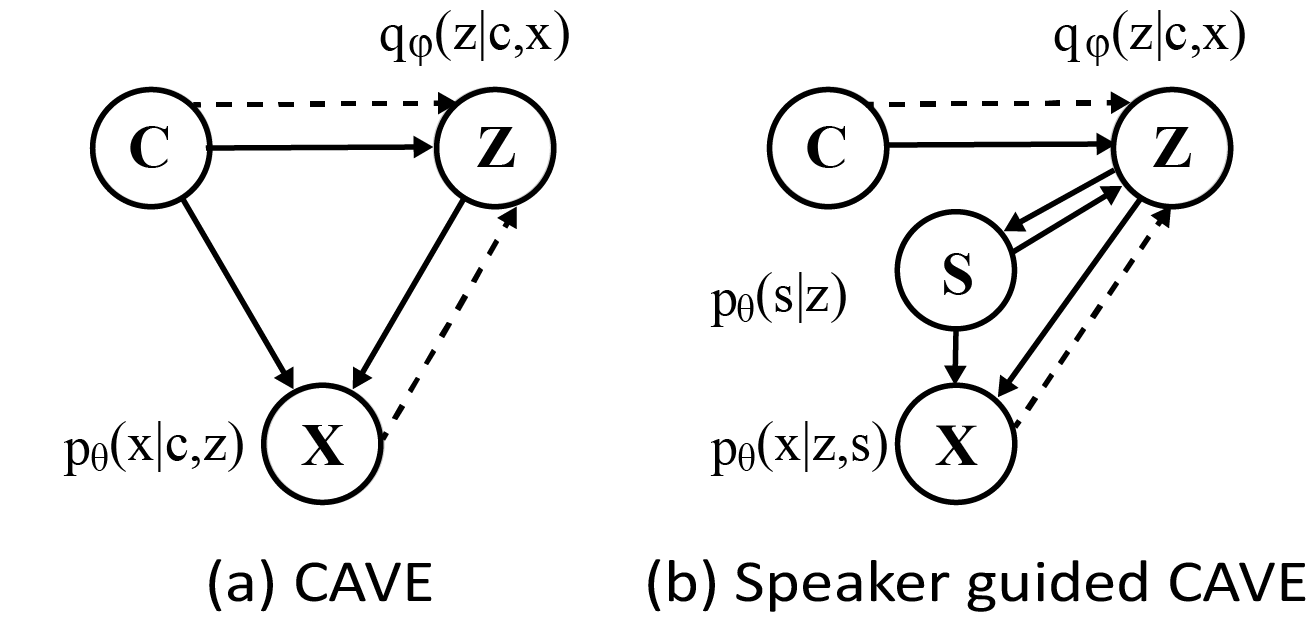}}
\caption{Graphical models of CVAE and Speaker-Guided CVAE~(sgCVAE).}
\end{figure}
VAEs~\cite{kingma2014stochastic} are one of the auto-encoder neural network architectures. The encoder network denotes the conditional distribution $q_{\phi}(Z|X)$ for the latent space variable $Z$ given input data $X$ and the decoder network means the distribution $p_{\theta}(X|Z)$ for the generated data $X$ given input vector $Z$. VAEs can learn the parameters from training data to make the $q_{\phi}(Z|X)$ consistent with the $p_{\theta}(Z|X) \propto p_{\theta}(X|Z)p(Z)$. The log marginal distribution of data $X$ can be lower-bounded by the evidence lower bound~(ELBO) as followed:
\begin{equation} \label{eq2}
      \mathcal{L}(\theta, \phi; X)  = \mathbb{E}_{q_{\phi}(Z|X)}[log(p_{\theta}(X|Z))] - KL[q_{\phi}(Z|X) || p(Z)],
\end{equation}
where the $KL(q_{\phi}|p_{\theta})$ denotes the Kullback-Leibler~(KL) divergence that the minimum value is zero when the two distributions are consistent. So we can minimize the KL divergence between $p_{\theta}(Z|X)$ and $q_{\phi}(Z|X)$ by maximizing the ELBO, which can be further reduced to formula Eq.\ref{eq2}. One of the typical way to modeling $p_{\theta}(Z|X)$ and $q_{\phi}(Z|X)$ is to assume normal distributions with a diagonal covariance matrix. As for the prior distribution $p(Z)$, it can be designed as a specific form according to the assumption that we need. The first term of Eq.\ref{eq2} is the autoencoder reconstruction error, and the second term is the KL divergence $q_{\phi}(Z|X)=\mathcal{N}(Z|\mu_{\phi}(x), \sigma_{\phi}(x))$ and  $p(Z)=\mathcal{N}(Z|0, I)$. Using a reparameterization $z=\mu_{\phi}(x) + \sigma_{\phi} \odot \epsilon, \epsilon \sim \mathcal{N}(\epsilon | 0, I)$, we can sample $\epsilon$ from its distribution to generate $z$ instead of sampling from $q_{\phi}(Z|X)$ directly, so that we can compute the gradient of the lower bound with respect to $\theta$.

Conditional VAEs are the extended version of VAEs and are trained to maximize the conditional log-likelihood of X given C. As shown in Fig. 1(a), the variational lower bound to be maximized becomes
\begin{equation}
    \begin{split}
        \mathcal{L}(\theta, \phi; X,C) 
        & = \mathbb{E}_{q_{\phi}(Z|X,C)}[\mathrm{log}(p_{\theta}(X|Z,C))] \\ & - KL[q_{\phi}(Z|X,C) || p_{\theta}(Z|C)]  \\
        & \leq \mathrm{log} p_{\theta}(X|C).
    \end{split}
\end{equation}
Be similar to VAEs, the prior distribution $p(Z|C)$ can be designed for the requirements. When we assume the latent Z is independent with C, the $p_{\theta}(Z|C) = p(Z)$. As for the neural network architecture, the differences between VAE and CVAE are that the encoder and decoder networks can take an auxiliary variable C as an additional input.


\subsection{Speaker-guided CVAE for Zero-shot Multi-speaker TTS}

Most multi-speaker TTS model consists of a phoneme encoder to encode the phoneme information to a latent variable $Z$, a decoder to generate the mel-spectrogram from $Z$ and a speaker representation added to $Z$ to control the timbre. Speaker representation can be a jointly-optimized looked-up embedding table or pretrained speaker verification system. 

Based on the multi-speaker model, we model the multi-speaker TTS using CVAE. The CVAE for multi-speaker TTS is composed of mel-spectrogram, phoneme, and the latent variable. To better fuse the features, we define $X$, $C$ simply as the output of mel encoder and phoneme encoder. The mel encoder utilizes the speaker encoder of AdaIN-VC~\cite{chou2019one}. The phoneme encoder and mel decoder are based on FastSpeech2~\cite{ren2020fastspeech}.

We define the conditional distribution $p(X,Z|C) = p_{\theta}(X|C,Z)p_{\theta}(Z|C)$,  and  $p_{\theta}(X|C,Z)$ and $p_{\theta}(Z|C)$ are modeled by the mel decoder network and the prior encoder network. As shown in Fig.1(a), we use a $q_{\phi}(Z|C, X)$ to approximate the true posterior $p_{\theta}(Z|C)$. But this assumption is counterintuitive, because the latent variable $Z$ should consist the speaker information if we want reconstruct the mel-spectrogram. So we propose the speaker-guided CVAE (as shown in Fig.1(b) and Fig.2(b)) that assume the prior of latent $Z$ is based on the speaker representation $S$, $p_{\theta}(Z|C)=p_{\theta}(Z|C,S)$, and then use $q_{\phi}(Z|C, X)$ to approximate $p_{\theta}(Z|C,S)$. We assume that the latent $Z$ contains the whole information of $C$, so $ p_{\theta}(X|C,Z,S) =  p_{\theta}(X|Z,S)$.
In speaker-guided CVAE, we modify the conditional distribution to $$p(X,Z|C, S) = p_{\theta}(X|Z,S)p_{\theta}(Z|C,S).$$ We hypothesise the latent variable $z$ follows the Gaussian distribution with a diagonal covariance matrix, so $q_{\phi}(Z|C, X)  \sim \mathcal{N}(\mu_{1}, \sigma_{1}^{2} \textbf{I}) $ and  $p_{\theta}(Z|C,S) \sim \mathcal{N}(\mu_{2}, \sigma_{2}^{2} \emph{\textbf{I}})$. Then the $q_{\phi}(Z|C, X)$, named recognition network, and the $p_{\theta}(Z|C, S)$ named prior network are represented as:\begin{equation}
    \begin{split}
    & \left[\begin{array}{cc}
         \mu_{1}  \\
         \ln{log}\sigma^{2}_{1}
    \end{array}\right] = \textbf{MLP}_{q_{\phi}}(\left[\begin{array}{cc}
         X  \\
         C
    \end{array}\right]), \\
   &  \left[\begin{array}{cc}
         \mu_{2}  \\
         \ln{log}\sigma^{2}_{2}
    \end{array}\right] = \textbf{MLP}_{p_{\theta}}(\left[\begin{array}{cc}
         C  \\
         S
    \end{array}\right]).
    \end{split}
\end{equation}The reparametrization trick is used to sample $Z_{1}, Z_{2}$ from the distribution $\mathcal{N}(\mu_{1}, \sigma_{1}^{2} \emph{\textbf{I}})$ and $\mathcal{N}(\mu_{2}, \sigma_{2}^{2} \emph{\textbf{I}})$. Then, we can generate mel-spectrogram $X$ based on $p_{\theta}(X|Z,S)$, which is represent by mel decoder network. Additionally, We utilize a network to predict $ \widehat{S} = \textbf{MLP}_{S}(Z)$. In the inference step, the speaker embedding $S$ will be replaced by predicted $\widehat{S}$. Finally, we can use Hifi-GAN to generate voice~\cite{kong2020hifi}. 
\begin{figure}[htb]
\begin{minipage}[b]{1\linewidth}
  \centering
  \centerline{\includegraphics[width=8.5cm]{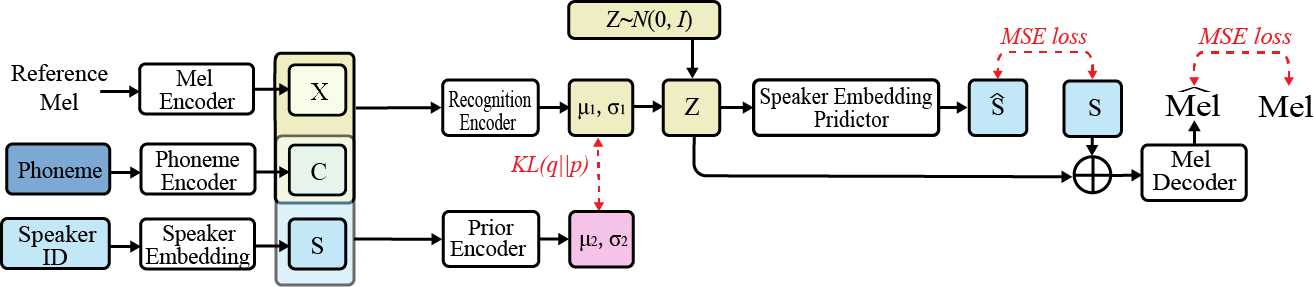}}
  \centerline{(a) Training}\medskip
\end{minipage}
\begin{minipage}[b]{1\linewidth}
  \centering
  \centerline{\includegraphics[width=7.5cm]{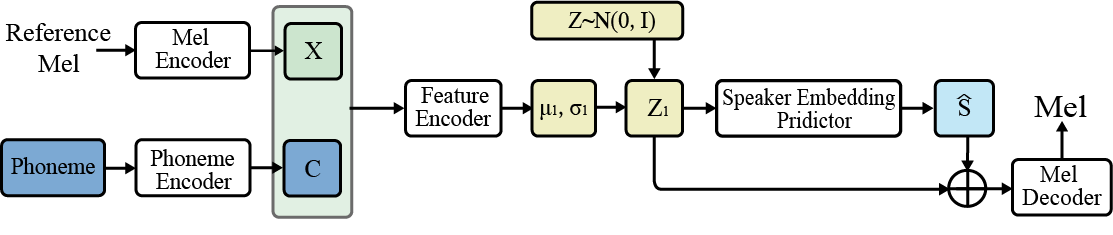}}
  \centerline{(b) Inference}\medskip
\end{minipage}
\caption{Neural network architectures of nnSpeech. }
\label{fig:fig2}
\end{figure}
\subsection{Loss Functions}
 nnSpeech model is trained by maximizing the modified ELBO:
 \begin{equation}
     \begin{split}
        \mathcal{L}(\theta, \phi; X,C,S) 
        & = \mathbb{E}_{q_{\phi}(Z|X,C)}[log(p_{\theta}(X|Z,S))] 
        \\ &+\mathbb{E}_{q_{\phi}(Z|X,C)}[log(p_{\theta}(S|Z))] 
        \\ & - KL[q_{\phi}(Z|X,C) || p_{\theta}(Z|C,S)],
     \end{split}
 \end{equation}
 where the first two terms are the expectation of generated mel-spectrogram and speaker embedding. Maximizing them means minimizing the mean square error of the estimations and the coefficient of the third term, KL divergence, is negative. Therefore, Maximizing the ELBO can be implemented by minimizing three loss functions, mel loss, speaker loss, and KL loss. The mel loss and speaker loss are defined as: \begin{equation}
     \begin{split}
& \mathcal{L}_{mel} = \mathbb{E}[\lvert \lvert X- \widehat{X} \rvert \rvert_{2}],\\
& \mathcal{L}_{spk} = \mathbb{E}[\lvert \lvert S- \widehat{S}\rvert \rvert_{2}],\end{split}
 \end{equation}
where the X means mel-spectrogram and S denotes speaker embedding vector. The KL loss is calculated by
\begin{equation}
    \begin{split}
\mathcal{L}_{kl} & = \frac{1}{2} \mathrm{log} \frac{|\Sigma_{2}|}{|\Sigma_{1}|} -n +      tr(\Sigma_{2}^{-1}\Sigma_{1}) \\ & + (\mu_{1}-\mu_{2})^{T}\Sigma^{-1}_{2}(\mu_{1}-\mu_{2}),
    \end{split}
\end{equation}
 where the $\mu_{1}, \Sigma_{1}$ and $\mu_{2}, \Sigma_{2}$ denotes the Gaussian distribution $p_{\theta}$ and $q_{\phi}$. Then the total loss is:
 \begin{equation} 
       \mathcal{L} 
       = \alpha*\mathcal{L}_{mel}+\beta*\mathcal{L}_{spk} 
       +\gamma*\mathcal{L}_{kl} 
 \end{equation}
$\alpha$ and $\beta$ are set as 1, and $\gamma$ is the hyper-parameter of our model that will be analyzed in the following experiments. Besides, our model has duration loss, energy loss, and pitch loss like Fastspeech2. 
 


\section{EXPERIMENTS}
\label{sec:typestyle}

\subsection{Experimental Setup}
We trained our model on a multi-speaker English corpus~(903 speakers, 191.29 hours, a subset 'train-clean-360' of the whole LibriTTS~\cite{zen2019libritts}),  and a multi-speaker Mandarin corpus~(218 speakers, 85 hours, AISHELL-3~\cite{shi2021aishell3}).
We randomly chose several male and female speakers from both the LibriTTS and AISHELL3 for evaluation. The only single speaker dataset LJSpeech~\cite{ito2017lj} was used to do a cross-dataset evaluation.

The speech data were resampled to 22050Hz and then extracted the mel-spectrogram using 256 hop size and 1024 window size for all datasets. We also calculated the duration, pitch, and energy for variance adaptor modular as the same as FastSpeech2~\cite{ren2020fastspeech}. 

In the training stage, we first trained the model without speaker embedding predictor module 100,000 steps, and then trained the whole model without speaker embedding module 500,000 steps on NVIDIA V100 GPU with the batch size 16. We used the Adam optimizer with $\beta_{1}=0.9$, $\beta_{2}=0.98$, $\epsilon = 10^{-9}$. In the testing stage, one speech was randomly selected as the reference for each speaker. 

\begin{table*}[h]
  \centering
  \fontsize{8}{9}\selectfont
  \caption{Subjective Comparison of the fine-tune based TTS methods}
  \label{Comparison1}
\begin{tabular}{ccccccccc}
    \toprule   
    \multicolumn{1}{c}{\multirow{2}{*}{Methods}} & \multicolumn{1}{c}{\multirow{2}{*}{Params/spk}} & \multicolumn{1}{c}{\multirow{2}{*}{Time/spk}} & \multicolumn{3}{c}{\textbf{MOS}} & \multicolumn{3}{c}{\textbf{VSS}} \\
    \cmidrule(lr){4-6} \cmidrule(lr){7-9}
    \multicolumn{1}{c}{}& \multicolumn{1}{c}{}   & \multicolumn{1}{c}{}     & \multicolumn{1}{c}{LibriTTS}       & AISHELL3    & LJSpeech    & LibriTTS   & AISHELL3     & LJSpeech   \\ 
    \midrule   
    GT      & --  & --   &  4.73 $\pm$ 0.49 & 4.60 $\pm$ 0.60 & 4.75 $\pm$ 0.60 & -- & -- & --  \\
    GT+Vocoder  & -- & --  & 4.48 $\pm$ 0.69  & 4.52 $\pm$ 0.68 & 4.71 $\pm$ 0.54  & 4.84 $\pm$ 0.68    & 4.79 $\pm$ 0.57 & 4.92 $\pm$ 0.43  \\
    \midrule 
    Finetune spk emb  & 256  & 1410 $\pm$ 396  
    & 3.45 $\pm$ 0.92 & 4.10 $\pm$ 0.87  & 3.33 $\pm$ 0.90  
    & 3.56 $\pm$ 1.03  & 3.91 $\pm$ 0.97 & 2.88 $\pm$ 1.00  \\
    Finetune decoder  & 14 M   & 1690 $\pm$ 526   
    & \textbf{3.48} $\pm$ \textbf{0.88}  & \textbf{4.25} $\pm$ \textbf{0.83} & \textbf{3.39} $\pm$ \textbf{0.91}  
    & \textbf{4.03} $\pm$ \textbf{0.98}  &\textbf{4.68} $\pm$ \textbf{0.77} & \textbf{4.00} $\pm$ \textbf{0.87} \\
    \midrule   

    X-vector~\cite{snyder2018x} & --   & --   
&3.27 $\pm$ 0.85 & 3.95 $\pm$ 0.84 & 3.03 $\pm$ 0.94
& 3.30 $\pm$ 0.81  &3.86 $\pm$ 0.82 & 2.21 $\pm$ 0.88 \\
 StyleSpeech~\cite{min2021meta} & --   & --  
 & 3.33 $\pm$ 0.91 & 4.07 $\pm$ 0.91 & \textbf{3.37} $\pm$ \textbf{0.90}  
 & 3.36 $\pm$ 0.98 & 3.97 $\pm$ 0.76 & \textbf{3.24} $\pm$ \textbf{0.78} \\
 
  \textbf{nn-Speech} & -- & -- 
  & \textbf{3.42} $\pm$ \textbf{0.92}   & \textbf{4.13} $\pm$ \textbf{0.86}  & 3.13 $\pm$ 0.98  
  & \textbf{3.46} $\pm$ \textbf{0.93}  & \textbf{4.17} $\pm$ \textbf{0.73} &2.65 $\pm$ 0.89 \\
   
    \bottomrule   
\end{tabular}
\end{table*}
\subsection{Evaluation}
\label{evaluation}

We evaluated our model both on subjective and objective tests. For subjective test, we focused on the adaption voice quality, which consists of the voice naturalness mean opinion score~(MOS), and the voice similarity score~(VSS), as shown in Table \ref{Comparison1}. Fourteen listeners judged each sentence, and the averaged MOS scores and VSS score were the final result. In addition, we also calculated the parameters that needs to be stored and the consumed time of fine-tune based methods. As for the adaption utterances, 20 sentences were used for retraining 2,000 steps for fine-tune-based methods, and one same reference utterance was selected randomly for the one-shot multi-speaker methods. Besides, we calculated the mel-cepstral distortion~(MCD) score as the objective test.

We compared our method with fine-tuning the different parts of a multi-speaker TTS model. It could be found that fine-tuning the decoder could achieve a much better performance than fine-tuning the speaker embedding and zero-shot methods. For zero-shot multi-speaker methods, we compared our model with two Fastspeech2-based methods: 1). Using the x-vector~\cite{snyder2018x} as the speaker embedding in Fastspeech2; 2). StyleSpeech~\cite{min2021meta}. Our method outperformed the x-vector in three datasets and got better performance than StyleSpeech both on the MOS and VSS on AISHELL3 and LibriTTS. However, our model performed poorly under cross-dataset LJspeech. Both fine-tuning decoder and using the style adaptive layer normalization~(StyleSpeech) perform better than fine-tuning speaker embedding and using x-vector.

\begin{table}[h]
  \centering
  \fontsize{8}{9}\selectfont
  \caption{Objective Comparison of the fine-tune based methods~(MCD)}
  \label{Comparison2}
\begin{tabular}{ccccccc}
    \toprule   
    \multicolumn{1}{c}{\multirow{1}{*}{Metric}} &  \multicolumn{1}{c}{LibriTTS}  & \multicolumn{1}{c}{AISHELL3} &\multicolumn{1}{c}{LJSpeech}\\
    \midrule   
    Finetune spk emb  & 7.55 $\pm$ 0.64 & 8.29 $\pm$ 0.65 & 7.65 $\pm$ 0.66  \\
    Finetune decoder & \textbf{7.18} $\pm$ \textbf{0.57}  & \textbf{7.74} $\pm$ \textbf{0.57}  & \textbf{7.05} $\pm$ \textbf{0.50}  \\
    \midrule
    X-vector & 7.95 $\pm$ 0.65  &  8.89 $\pm$ 0.54  & 8.73 $\pm$ 0.68   \\
    StyleSpeech & 7.82 $\pm$ 0.68    &8.64 $\pm$ 0.67  & \textbf{7.38} $\pm$ \textbf{0.51} \\
    \textbf{nnSpeech}  & \textbf{7.44} $\pm$ \textbf{0.59} & \textbf{8.38} $\pm$ \textbf{0.53} & 8.27 $\pm$ 0.54  \\
    \bottomrule   
\end{tabular}
\end{table}

\subsection{Method Analysis}

In this section, we first studied the voice quality with different adaption data. Next, we analyzed the hyper-parameter $\gamma$, the weight of KL divergence loss. In addition, we conduct the ablation studies.

\textbf{MCD under vary adaption data}. We studied the MCD score with different adaption data on the AISHELL3. For our model, we replaced the output of mel encoder $X$ and predicted $\hat{S}$ by their averaged vectors. It can be seen that the MCD score decreases slightly with the increase of the adaption data, as shown in Fig.3. Our method could get better performance when using four adaption utterances, which proves that our method could extracts the speaker information from a few adaption data. 

\begin{figure}[htb] \label{fig: fig3}
  \centering
  \centerline{\includegraphics[width=9cm]{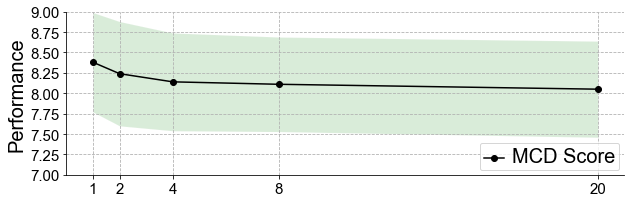}}
\caption{Performance with varying number of adaption voices.}
\end{figure}

\textbf{Hyper-parameter analysis.} As shown in Table \ref{Comparison3}, we analyzed the hyper-parameter $\gamma$ on the dataset AISHELL3, and the model got the best performance when the $\gamma$ was set as 0.0005 compared with the set as 0.005 and 0.05. That is because the KL divergence is always large than MSE loss, such as $\mathcal{L}_{mel}$ and $\mathcal{L}_{spk}$. So the model will be hard to learn something from other losses if the weight of KL loss is large.

\begin{table}[h]
  \centering
  \caption{The MCD with different $\gamma$ setting on AISHELL3}
  \label{Comparison3}
\begin{tabular}{cccc}
    \toprule   
    \multicolumn{1}{c}{\multirow{1}{*}{Metric}} &  \multicolumn{1}{c}{$\gamma$=0.05}  & \multicolumn{1}{c}{$\gamma$=0.005} &\multicolumn{1}{c}{$\gamma$=0.0005}\\
    \midrule   

    MCD & 8.90 $\pm$ 0.49  & 8.50 $\pm$ 0.51  & \textbf{8.38} $\pm$ \textbf{0.53} \\
    \bottomrule   
\end{tabular}
\end{table}

\begin{table}[h]
  \centering
  \caption{The ablation experiment on AISHELL3}
  \label{Comparison4}
\begin{tabular}{ccccccc}
    \toprule   
    \multicolumn{1}{c}{\multirow{1}{*}{Metric}} &  \multicolumn{1}{c}{Content add}  & \multicolumn{1}{c}{w/o spk pred} & \multicolumn{1}{c}{nnSpeech} \\
    \midrule   
  MCD & 8.83 $\pm$ 0.45 & 8.41 $\pm$0.63 & \textbf{8.38} $\pm$ \textbf{0.53}\\
    \bottomrule   
\end{tabular}
\end{table}

\textbf{Analyses on the model architecture.} We further compared the model architecture with two other settings: 1). Add the content information $C$ to the latent variable $Z$, so our model will become the standard CVAE; 2). remove the speaker embedding predictor module and not add the predicted speaker embedding $C$ to the latent $Z$. As shown in Table \ref{Comparison4}, we found that both settings get the higher MCD than our model. For setting 1, we believe that the latent variable $Z$ involves the content information and speaker information, so adding with the content $C$ again is not necessary and maybe cover up the speaker information. For setting 2, the result proves that the speaker predictor module is helpful.

\section{CONCLUSION}
\label{sec:majhead}
We proposed the speaker-guided CVAE for zero-shot multi-speaker TTS named nnSpeech. Our method could improve the generalization of the zero-shot TTS model. It performs excellent when only using one adaption voice both in the English and Chinese corpus. We have done some model analysis to find the best hyper-parameter and proves the our model is robust and each module is helpful. However, our model performs poorly on the cross-dataset, and there is still a gap between zero-shot methods and fine-tune methods. For future work, we will focus on the cross-dataset problem. The result in section \ref{evaluation} shows using adaptive layer normalization to modulate the decoder is a possible solution.

\section{acknowledgement}
\thanks{This paper is supported by the Key Research and Development Program of Guangdong Province No. 2021B0101400003 and the National Key Research and Development Program of China under grant No. 2018YFB0204403.}

\vfill\pagebreak

\bibliographystyle{IEEEbib}
\bibliography{strings,refs}

\end{document}